\documentclass{acm_proc_article-sp}

\usepackage{algorithm}
\usepackage{algorithmic}
\usepackage{acronym}
\usepackage{amsmath}
\usepackage{amssymb}
\usepackage{balance}
\usepackage{cite}
\usepackage{enumerate}
\usepackage{graphicx}
\usepackage{subfigure}
\usepackage{textcomp}
\usepackage{verbatim}
\usepackage[usenames]{color}
\usepackage{url}
\usepackage{etoolbox}
\usepackage{multirow}
\usepackage{tabularx}

\patchcmd{\maketitle}{\@copyrightspace}{}{}{}
\makeatother

\begin{document}

\clubpenalty=10000
\widowpenalty=10000
\makeatother
\title{2DR: Towards Fine-Grained 2-D RFID Touch Sensing}

\vspace{-4cm}

\author{
      \centerline{Shilin Zhu*, Yilong Li*}\\
      \centerline{($^*$Co-primary authors)}\\
      \affaddr{University of California San Diego, University of Wisconsin-Madison}\\
      \email{shz338@eng.ucsd.edu, yli758@wisc.edu}
}

\maketitle

\begin{abstract}
	In this paper, we introduce \textit{2DR}, a single RFID tag which can seamlessly sense two-dimensional human touch using off-the-shelf RFID readers. Instead of using a two-dimensional tag array to sense human finger touch on a surface, \textit{2DR} only uses one or two RFID chip(s), which reduces the manufacturing cost and makes the tag more suitable for printing on flexible materials. The key idea behind \textit{2DR} is to design a custom-shape antenna and classify human finger touch based on unique phase information using statistical learning. We printed \textit{2DR} tag on FR-4 substrate and use off-the-shelf UHF-RFID readers (FCC frequency band) to sense different touch activities. Experiments show great potential of our design. Moreover, {\color{black}{\textit{2DR} can be further extended to 3D by building stereoscopic model}}.
	
\end{abstract}

\section{Introduction}
\label{sec:intro}
We live in a world with emerging IoT applications. The smart and seamless interaction between human and IoT devices reshapes our understanding of intelligence, e.g., voice command, body language, and finger touch. To realize ubiquitous touch sensing, RFID (Radio Frequency IDentification) has become the most promising technology since it does not require battery to work so that tags can be placed everywhere around us. In the past few years people have proposed various ways to detect gestures such as swiping or touching, even used machine learning to classify interactions. However, these works mainly classify several patterns but do not realize fine-grained touch sensing over the tag surface. Recent works propose to use a two-dimensional RFID tag matrix, but each tag in matrix needs a chip which is not a flexible component. Moreover, they need to place enough tags to support continuous tracking, sacrificing size and cost.

In this paper, we design and build a single RFID tag, named \textit{2DR}, which can sense 2D finger touch on the tag surface. The key idea is that human finger can cause the phase pattern of RFID readings to change under different touch points. We design the tag such that each touching point leads to an unique phase pattern. However, we noticed a strong phase fluctuation caused by multi-path wireless environment. We suppress these noise by placing two chips on our tag, and evaluate phase difference between two chips, so that we can partially suppress the noise. We use traditional but very effective K-Nearest Neighbor (KNN) and Support Vector Machines (SVM) to recognize such pattern. The KNN detector is more attractive since \textit{2DR} model may need to fine-tune for different people before use. We print \textit{2DR} on FR-4 substrate with one and two RFID chips and use an off-the-shelf RFID reader in FCC band to demonstrate. \textit{2DR} has two main contributions to the community:

\textit{(i) Ubiquitous 2D touch sensing:} \textit{2DR} offers a promising direction for designing 2D or even 3D touch sensors using RFID tags and off-the-shelf readers. Our system using phase difference between two chips with pattern recognition provides an alternative way rather than using tag matrix.

\textit{(ii) Customized RFID tag antenna:} \textit{2DR} opens a world of designing low-cost RFID tags with only one or two chip(s) and special antenna shapes. We believe there will be more such customized tags in the near future for different sensing tasks in IoT era.

Although \textit{2DR} only uses phase information, we believe it can be further improved by adding other useful  features, e.g., from PHY layer, to enhance the robustness of \textit{2DR}, and by searching for optimal antenna shape design. 

\section{Related Work}
\label{sec:related}
\textbf{RFID ubiquitous sensing: }Researchers have built RFID systems for gesture recognition using phase and RSS signatures with tag matrix \cite{zou2017grfid}. Several papers also used RFID tags to sense environments, e.g., temperature \cite{amendola2016design} and track objects \cite{wei2016gyro}. RIO \cite{RIO} uses commercial RFID tag for finger tracking but can only identify 3 points in one dimension on a single tag because the phase change of touching a dipole antenna is symmetrical. \textit{2DR} uses a single antenna with one or two chip(s) to realize 2D touch sensing, outperforming traditional tag matrix solution in terms of cost, size, resolution and extensibility. 

\textbf{Pattern recognition using sensory data: }Pattern recognition and statistical learning have been proposed to classify people's daily activities \cite{ding2015femo, han2016cbid}. Due to the unpredictability of wireless channel, machine learning maybe the best way to guarantee a satisfying performance. For human touch sensing, the latency requirement is stringent. \textit{2DR} uses KNN to automatically classify touch points using phase difference in an unsupervised way without any labeling process. Results show that KNN has comparable performance compared with SVM trained using true labels.

\section{Design}
\label{sec:system}
RFID system works via backscatter communication. In a typical half-duplex RFID system, the reader sends query pulse and RFID tag backscatters sequential signals to the reader. UHF RFID system works on the FCC band from 902MHz to 928MHz in the United States. The limited bandwidth is a challenge for antenna design since the phase deviation outside frequency band can not be measured by off-the-shelf RFID reader. 

In \textit{2DR}, we design UHF RFID tag to enable 2-D touch sensing applications. The core idea is to produce unique touching features on custom-designed RFID antennas that can be measured by a RFID reader. We first design a double-spiral antenna with single chip and leverage phase information to calculate the coordinate of touching points in two dimensions on the tag. To overcome environmental noise and multi-path effect, we further design a double-spiral antenna with two chips. The phase value difference between two chips is a more robust metric, according to both our simulator and real measurement.

\subsection{Single-Chip \textbf{\textit{2DR}} Tag}

We tried multiple antenna shapes before we figure out the optimal design for 2-D touch sensing. A key challenge of touch sensing using RFID tags is variation of different touching behavior, e.g.\ the impedance of antenna will change differently when people touch in different way. Furthermore, different people also have different skin impedance. A feasible way is to use electromagnetic simulation before we fabricate actual RFID tags. We first simulated spiral antenna and dipole antenna to see if there is any feature when human finger has contact with metal sheet part. However, conductivity of skin is not stable since human body hydration percentage changes through time. We build a finger model with a cylinder of \textit{1cm} diameter and set a dynamic range of dielectric constant. According to simulated S-parameter (which describes the electrical behavior such as transmission and reflection), {\color{black}we eventually choose spiral shape antenna instead of dipole antenna. Many commercial RFID tags are designed as strip shape which limits the capability of sensing 2-D touch event.}

\begin{figure}[t]
    \centering
    \includegraphics[width=0.5\textwidth]{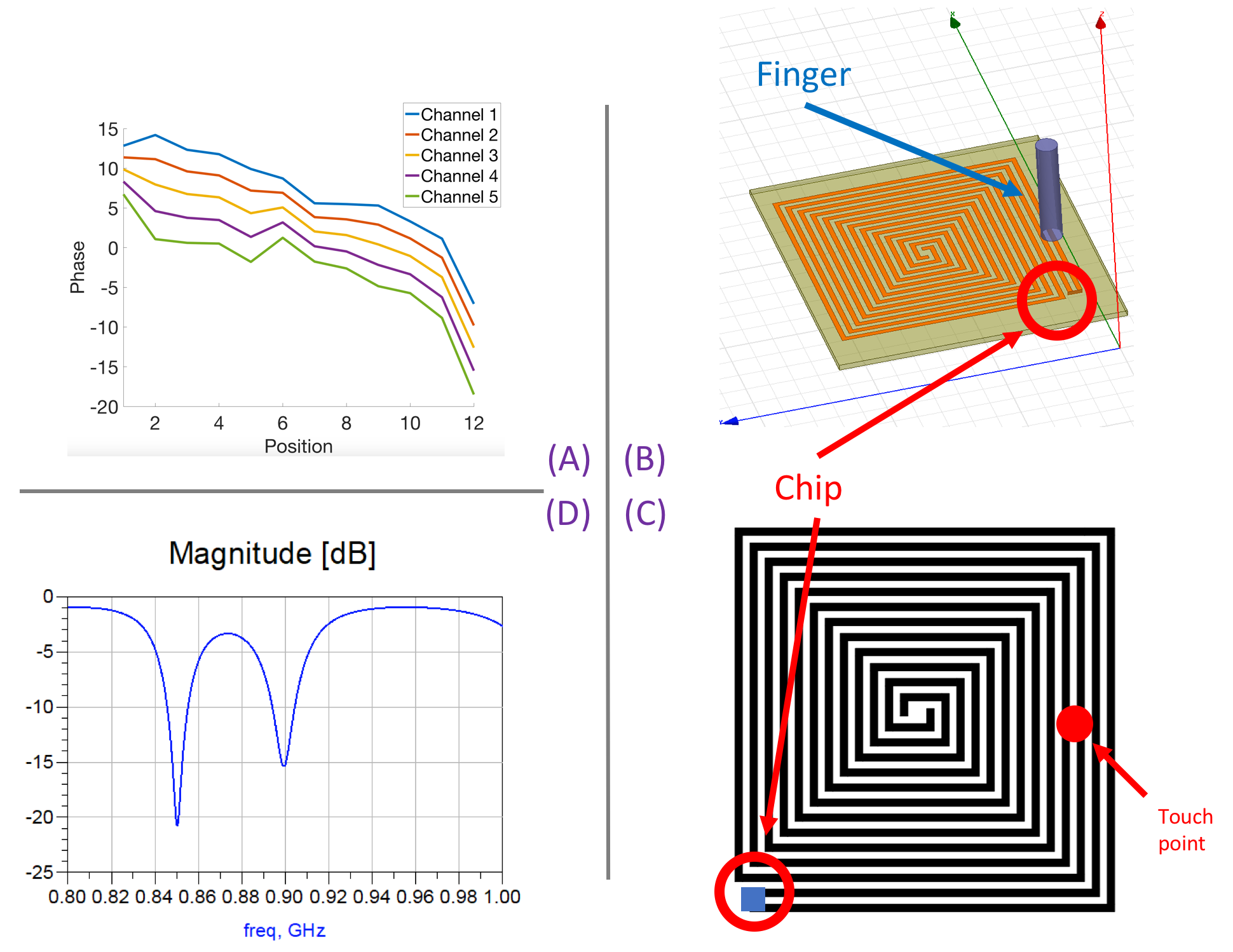}
    \caption{The design of single-chip tag. (A) Phase value with touch point. (B)(C): Antenna simulation interface and chip assignment. (D) Frequency response of tag.}
    \label{fig:single}
    \vspace{-0.1in}
\end{figure}

The material of substrate is also essential for \textit{2DR}. Existing industrial tag manufacturers always fabricate inlay on high quality printed polypropylene substrate via thermal transfer. But it is not suitable for experimental usage and fast fabrication. Instead we fabricated our RFID antennas on FR-4 substrate which is a type of glass epoxy, and is one of the most popular materials for fabricating circuit boards. Moreover, the thickness of the substrate is also an important factor for tag antenna design since it determines how electromagnetic wave propagates in the media of substrate. We choose the FR-4 substrate with a thickness of 1.5875mm for convenience of experiments, {\color{black}it has the dielectric constant around 4.4-4.6}. In our single-chip double-spiral tag, there are two same staggered spirals, each having size of 45.5mm $\times$ 45.5mm. The chip controls RFID system and can reduce noise to improve robustness of RFID communication. In this work, we pick \textit{Monza r6} produced by Impinj. Fig.\ref{fig:single} shows the antenna design, simulation model, frequency response and phase value of our single-chip tag. According to our simulation, the phase value changes decreasingly when the simulated 'finger' slides from the outer spiral into the inner one. However, simulation indicates an ideal scenario without considering the complex wireless environment and the interaction between skins and conductor is more complicated in practical scenario. 

\subsection{Double-Chip \textbf{\textit{2DR}} Tag}
\begin{figure}[bth]
    \centering
    \includegraphics[width=0.5\textwidth]{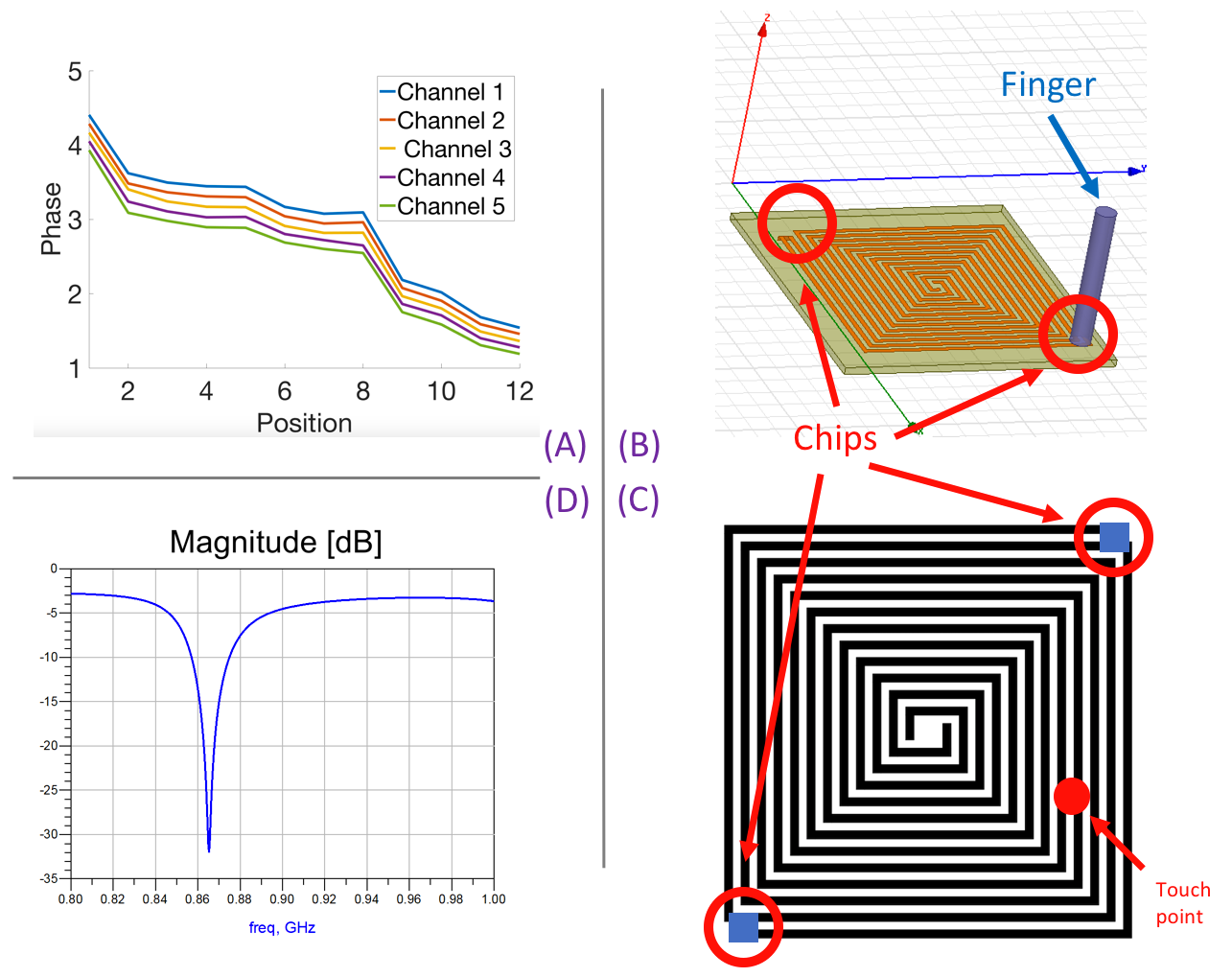}
    \caption{The design of double-chip tag.}
    \label{fig:double}
    \vspace{-0.1in}
\end{figure}

Although our single-chip tag looks good enough, the phase does not follow the simulation curve very well due to environmental noise and multi-path effect in real world. To deal with this problem, we design a double-chip tag, which uses the same double-spiral antenna as the single-chip tag but has two feeding ports (Fig.\ref{fig:double}). We install two RFID transponders to these two ports and consider the electromagnetic interference between them. Even each part of tag can work properly itself, the frequency response can be different when putting them together, and we need to guarantee both chips work in FCC band. Suppose the phase changes of a single chip caused by touching and environment interference are $\phi_{t}$ and $\phi_{e}$, then the phase difference $\Delta \phi$ between two chips can partially cancel $\phi_{e}$ (although not perfectly) and retain the effect caused by touching. According to our simulation,  $\Delta \phi$ changes decreasingly when we slide the antenna along diagonal. The phase difference of these two chips reflects the relative distance from touching point and can be used to co-locate the finger.

\subsection{Classify and Recognize Touching Points}
Even if we use double-chip \textit{2DR} tag, the environment still has non-negligible impact on the recorded phase values. To overcome this, we leverage learning approach to generate a robust model with recorded phase features. Specifically, we tried kernel SVM, fine-KNN and subspace-KNN:

\textbf{Kernel SVM: }SVM maximizes the margin of classifier's decision boundary, using kernel (e.g., RBF) to map to a non-linear feature space. It outperforms traditional perceptron method by searching for the best possible non-linear decision boundary in high-dimensional space.

\textbf{KNN: }KNN clusters data points using $K$ nearest neighbors and iterate until the clusters become stable. Fine-KNN uses only one neighbor. Subspace-KNN uses random subspace classification, which can ensemble multiple different models to increase the accuracy and robustness of classification.

\section{Evaluation}
\label{sec:evaluate}

\begin{figure}[bth]
    \centering
    \includegraphics[width=0.3\textwidth]{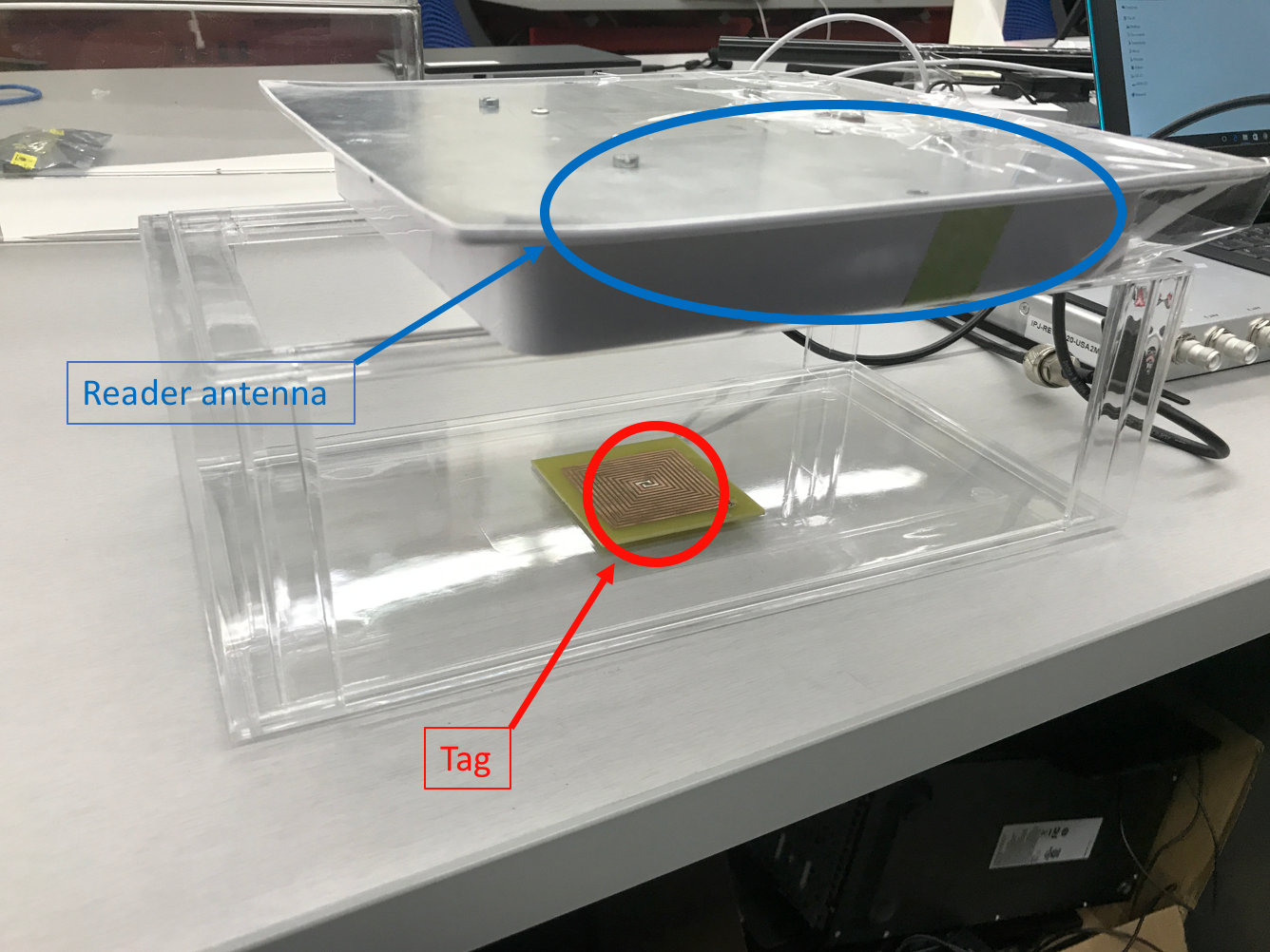}
    \caption{Experiment setup.}
    \label{fig:setup}
    \vspace{-0.1in}
\end{figure}

In experimental setup, we use \textit{SpeedWay RFID reader R420} for measurement which connects to a patch antenna on FCC band. The \textit{2DR} tag is deployed as shown in Fig.\ref{fig:setup}. We first test the single-chip tag but the resulting phase values do not follow the simulation very well, which is under our expectation due to changeability of noise. One possible reason is the existence of hand blocking where human hand blocks the wireless signal in different way each time. Environmental interference and multi-path effect also affect measured phase value. Simply using one transponder is not robust and it inspires us to favor our double-chip tag design.

\begin{figure}[t]
	\centering
	\begin{minipage}[t]{0.48\linewidth}
		\includegraphics[width=\linewidth]{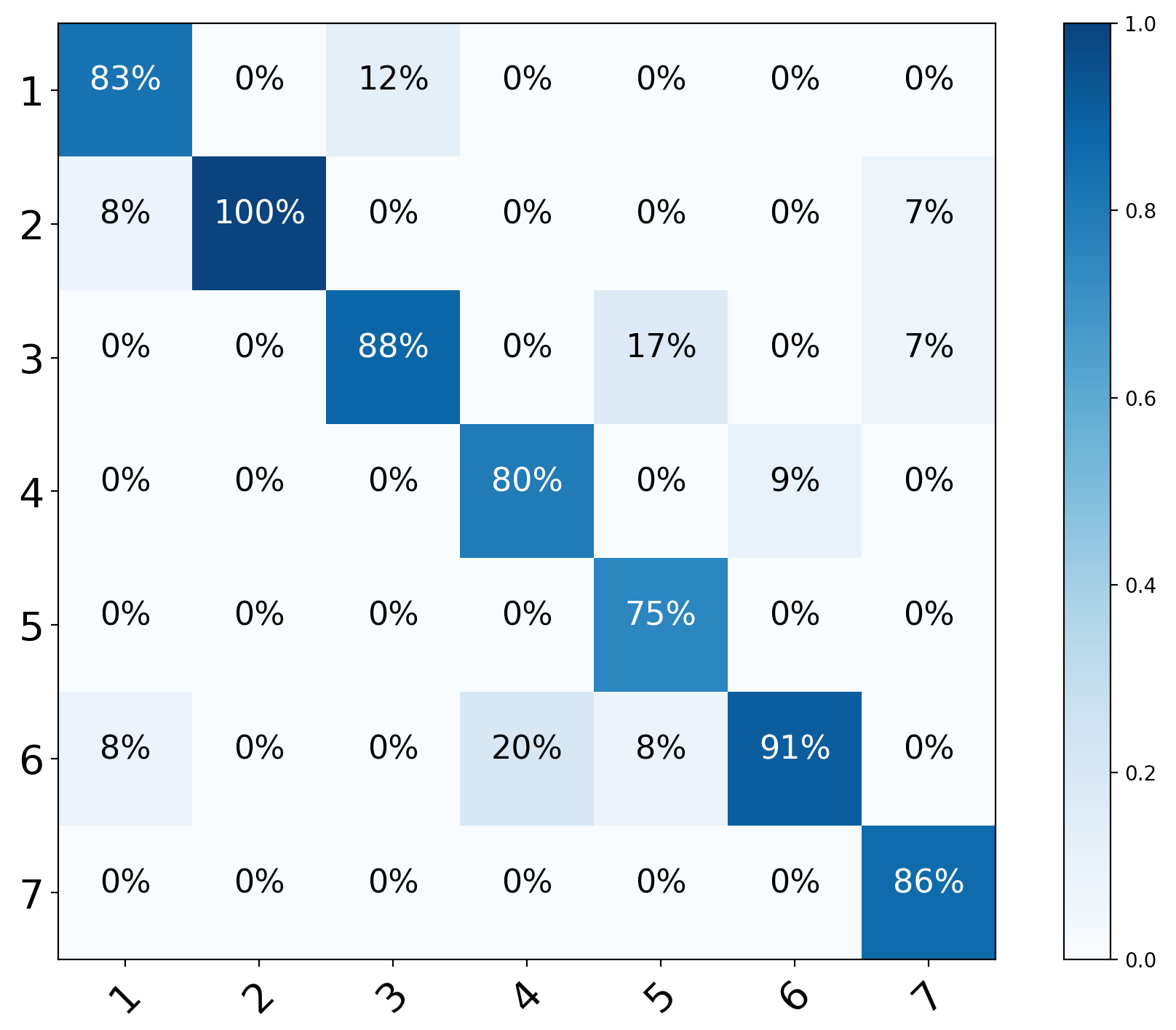}
		
	\end{minipage}
	\hspace{1mm}
	\begin{minipage}[t]{0.48\linewidth}
		\includegraphics[width=\linewidth]{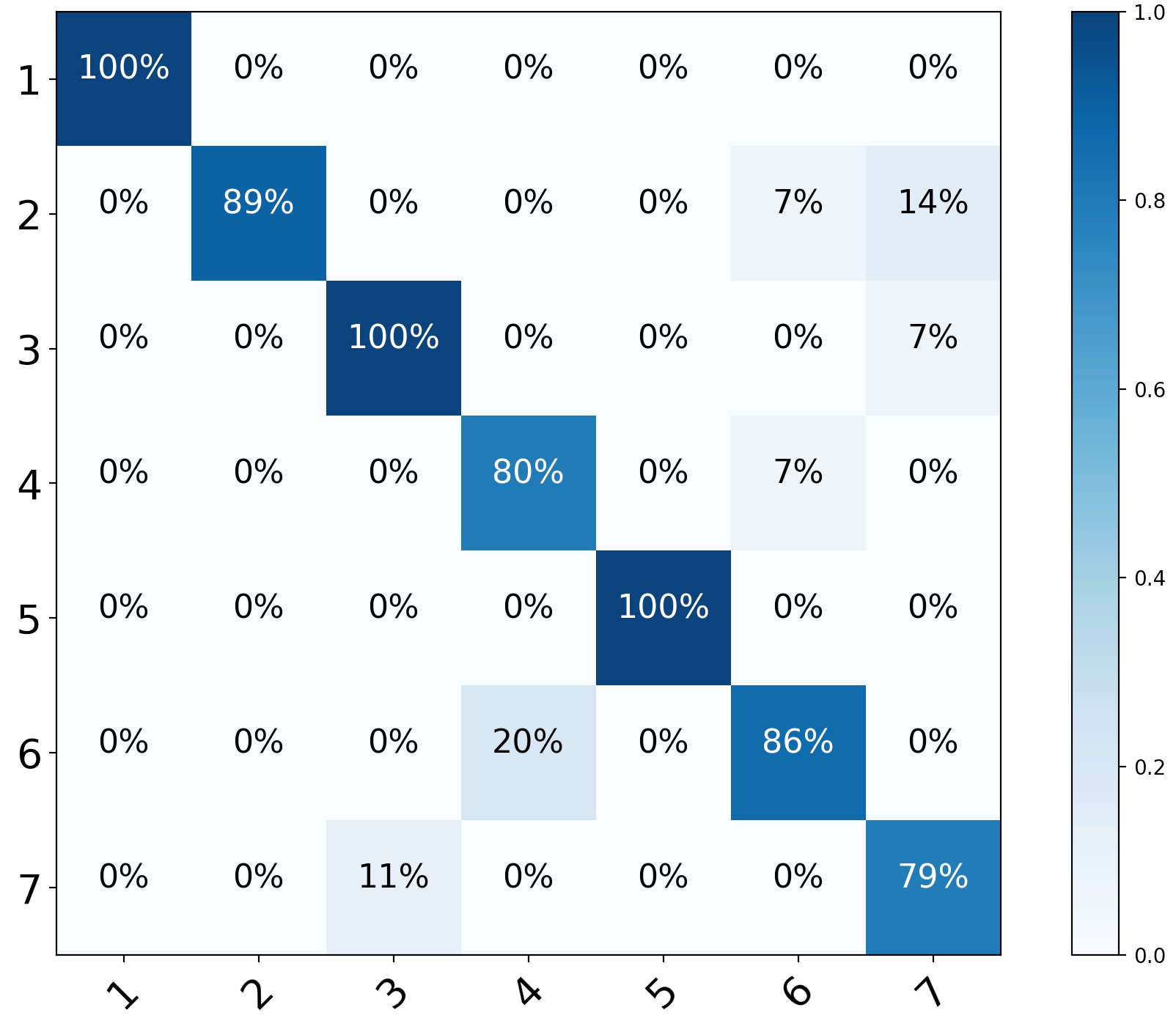}
		
	\end{minipage}
	\caption{(L)SVM and (R)Subspace-KNN confusion matrices.}
	\label{fig:conf_mat}
	\vspace{-0.1in}
\end{figure}

In order to test our double-chip tag, we touch tag surface and calculate the phase difference between two transponders for each point. Our experiments show that phase differences are distributed in 2D and generate an oblique plane, which serves as good feature to help us locate the touching point. To classify different points and overcome noise of data, we implement quadratic-SVM, fine-KNN and subspace-KNN and train the model using phase difference data recorded in 7 different touching points on our double-chip tag. The sample length, training time and accuracy are listed in Table \ref{tab:learn} and confusion matrices are shown in Fig. \ref{fig:conf_mat}. Results have shown satisfying performance of \textit{2DR}. Putting \textit{2DR} tag in a new wireless environment requires re-training process to be executed. Fortunately, Table \ref{tab:learn} shows that training can be very fast using SVM and KNN.

\begin{table}[]
\begin{tabular}{|c|c|c|c|}
\hline
Model         & \multicolumn{1}{l|}{Quad-SVM} & \multicolumn{1}{l|}{Subspace-KNN} & \multicolumn{1}{l|}{Fine-KNN} \\ \hline
\#Points      & 50                                 & 50                                & 50                            \\ \hline
Training  & 3.4995s                            & 2.7512s                           & 0.69095s                      \\ \hline
Accuracy      & 85.5\%                             & 89.5\%                            & 82.9\%                        \\ \hline
\end{tabular}
\caption{Classify 7 touching points with learning.}
\label{tab:learn}
\vspace{-0.1in}
\end{table}

\section{Summary}
\label{sec:dis_con}
\textit{Customized RFID antenna: }\textit{2DR} conveys a message: leveraging the electromagnetic properties of antenna with specific shape has greater potential compared with traditional way of stacking off-the-shelf tags. This is because application-driven antennas can serve better in interaction between human and IoT devices. \textit{2DR} has advantages in terms of cost, size, sensing resolution and printing flexibility.

\textit{Environmental interference: }Although \textit{2DR} uses the phase difference of two chips to increase robustness, it still lacks ability to fully overcome environmental noise. This is because phase information cannot fully estimate the multi-path wireless channels although chip(s) can partially suppress multi-path effect, which is one of the common problems in RF sensing community. We believe multi-path free \textit{2DR} by acquiring other RF signals or designing a multi-path free antenna is an important next step.

{\color{black}{\textit{Flexibility: } Although we currently fabricate \textit{2DR} tags on FR-4 substrate of 1.575mm thickness which is hard to bend, our design is eaily transplanted on paper-like or flexible substrate to enable many ubiquitous computing applications. }}

To sum up, \textit{2DR} offers a promising way to realize  ubiquitous RFID sensing in IoT era. Although environmental interference is still an open issue, \textit{2DR} is able to realize fine-grained 2D touch sensing over the tag by characteristics of its antenna shape, and has huge advantage in terms of manufacturing cost, size and flexibility of printing.

\bibliographystyle{abbrv}
\small
\bibliography{main}
\clearpage

\end{document}